\documentclass{aa}
\usepackage{txfonts}
\usepackage{graphicx}
\usepackage{aalongtable}

\newcommand{\diff}{\mbox{${\rm d}$}}

\newcommand{\feh}{\mbox{\rm [{\rm Fe}/{\rm H}]}}
\newcommand{\mh}{\mbox{\rm [{\rm M}/{\rm H}]}}

\newcommand{\Teff}{\mbox{$T_{\rm eff}$}}

\newcommand{\comment}[1]{}
\newcommand{\beq}{\begin{equation}}
\newcommand{\eeq}{\end{equation}}
\newcommand{\beqa}{\begin{eqnarray}}
\newcommand{\eeqa}{\end{eqnarray}}
        \def\smallskip{\vskip 2pt}

\begin{document}

\title{On the effect of helium enhancement on bolometric corrections 
and $\Teff$-colour relations}
\subtitle{}

\author{L. Girardi$^{1}$, F. Castelli$^{2}$, G. Bertelli$^{1}$, E. Nasi
$^{1}$}
\institute{
 Osservatorio Astronomico di Padova, 
	Vicolo dell'Osservatorio 5, I-35122 Padova, Italy  \and
 Osservatorio Astronomico di Trieste, 
	Via G.B.\ Tiepolo 11, I-34131 Trieste, Italy 
}

\offprints{L\'eo Girardi \\ e-mail: leo.girardi@oapd.inaf.it} 

\date{To appear in Astronomy \& Astrophysics}

\abstract{
We evaluate the effects that variations in He content have on
bolometric corrections and \Teff-colour relations. To this aim, we
compute ATLAS9 model atmospheres and spectral energy distributions for
effective temperatures ranging from 3500~K to 40000~K for dwarfs and
from 3500~K to 8000 K~for giants, considering both ``He-non enhanced''
and ``He-enhanced'' compositions. The considered variations in He
content are of $\Delta Y = +0.1$ and $+0.2$ for the metallicity
$\mh=+0.5$ and $\Delta Y = +0.1$ for $\mh=-0.5$ and $-1.5$. Then,
synthetic photometry is performed in the $UBVRIJHK$ system. We
conclude that the changes in bolometric corrections, caused by the
adopted He-enhancements are in general too small (less than 0.01~mag),
for both dwarfs and giants, to be affecting present-day tables of
bolometric corrections at a significant level.  The only possible
exceptions are found for the $U$-band at \Teff\ between 4000~K and
8000~K, where $|\Delta BC_{U}|$ amounts to $\sim0.02$~mag, and for
\Teff\ equal to 3500~K, where $|\Delta BC_{S_\lambda}|$ values become
clearly much higher (up to 0.06~mag for passbands from $U$ to
$V$). However, even in the latter case the overall uncertainty caused
by variations in the He content may be not so significant, because the
ATLAS9 results are still approximative at their lowest temperature
limit.  }

\authorrunning{L. Girardi et al.}
\titlerunning{Synthetic photometry and He content}
\maketitle

\section{Introduction}
\label{intro}

Over the last years, many different problems have prompted the
computation of stellar evolutionary tracks for different values
of initial He content. Just to mention a few of them:

1) Evidence has been found for significant variations in He content in
some globular cluster like $\omega$~Cen, NGC~2808 and M~13 (Piotto et
al. 2005; Lee et al. 2005; D'Antona et al. 2005; Caloi \& D'Antona
2005). These variations suggest that the relationship between He and
metal content was not univocal during the first period of chemical
enrichemnt of the universe. A univocal relationship, instead, has so
far been assumed in most grids of stellar models applied to the study
of old stellar populations.

2) The primordial He content has been recently revised upward, from
$Y_{\rm p}\sim0.235$ to $Y_{\rm p}=0.248\pm 0.001$, after the WMAP
mission (Spergel et al. 2003, 2006). Many grids of models for population II
stars have been computed for $Y$ values lower than the WMAP one.

3) On one side the helium content in five Hyades binary systems
($Y=0.255$) is lower than expected from their supersolar metallicity,
pointing to a value $dY/dZ$ of the order of 1 (Lebreton et al, 2001),
whereas other observations either indicate higher values,
$dY/dZ\sim2-2.5$ (e.g Jimenez et al. 2003 from K dwarf stars in the
Hipparcos catalog; Peimbert et al. 2002 from extragalactic HII
regions), or fail to constrain it to a significant level ($dY/dZ=3\pm
2$, Pagel \& Portinari 1998). Grids of stellar models for population
synthesis (e.g. Bertelli et al. 1994; Girardi et al 2000), instead, in
general use high $dY/dZ$ values in order to fit both the primordial
and the solar initial He content.

These aspects have prompted us to start a large project for the
computation of stellar tracks covering a large region of the $Y-Z$
plane (Bertelli et al. in preparation). Once ready, these tracks will
allow us to model stellar populations at any intermediate $Y$, thus
taking into account the changes in lifetimes, luminosities and \Teff\
that follow from a varying $Y$.

However, before stellar evolutionary tracks and isochrones are
compared with observations, they have to be converted to magnitudes
and colours via bolometric corrections (BC) and colour-\Teff\
relations. The latter may also be affected by the changes in He
content, and the purpose of this paper is exactly to evaluate how
much. To do so, we first compute energy distributions for a few selected
chemical mixtures with different  $Y$ (Sect.~\ref{sec_spectra}), and then
perform synthetic photometry on them (Sect.~\ref{sec_synphot}). The
results, in terms of changes in BCs and colours, are discussed in
Sect.~\ref{sec_conclu}.

\section{Synthetic spectra for He-enhanced compositions}
\label{sec_spectra}

Small grids of ATLAS9 model atmospheres and energy distributions
(Castelli \& Kurucz 2003) were generated for different sets of
metallicities and enhanced helium contents.  For consistency reasons
between continuous and line opacities, new opacity distribution
functions (ODFs) were computed for each chemical composition having
enhanced helium abundance.  The DFSYNTHE code (Kurucz 2005; Castelli
2005) was used to this purpose.

The solar and scaled-solar abundances selected for this study are
based on the solar chemical composition from Grevesse \& Sauval
(1998).  They are the same ones used by Castelli \& Kurucz (2003) for
the ODFNEW grids of models and fluxes\footnote{The ODFNEW spectral
energy distributions from Castelli \& Kurucz (2003), {as well as
the He-enhanced ones presented in this paper,} are available at\\ {\tt
http://wwwuser.oat.ts.astro.it/castelli/grids.html}}. In terms of
fractional mass, the abundances are $X=0.735, Y=0.248, Z=0.0170$ for
the solar case. The solar and scaled-solar abundances will hereafter
be mentioned as the $\Delta Y=0$ case. Then, for 3 different values of
metal content, we have computed energy distributions for the following
mixtures:
\begin{itemize}
\item for $\mh=-1.5$: $\Delta Y=0$ and $\Delta Y=0.1$.  
\item for $\mh=-0.5$: $\Delta Y=0$ and $\Delta Y=0.1$.  
\item for $\mh=+0.5$: $\Delta Y=0$, $\Delta Y=0.1$ and $\Delta Y=0.2$.  
\end{itemize}
The $\mh=-1.5$ and $\mh=-0.5$ spectra aim to probe the effect of He at
globular cluster metallicities, whereas the $\mh=+0.5$ ones serve to
probe the potential effect at the supersolar metallicities found in
giant ellipticals, for which measurements of the He content do not
exist. The effects at solar metallicities are of course derivable by
interpolation between the $\mh=-0.5$ and $\mh=+0.5$ cases.

Then, for each one of these chemical mixtures, we compute energy
distributions for a sequence of dwarfs and giants at several \Teff\
values. They are:
\begin{itemize}
\item Dwarfs: with $\log g=4.5$, and for \Teff=3500, 4000, 5000, 6000, 
8000, 12000, 20000, and 40000 K.
\item Giants: with $\log g=1.5$, and for \Teff=3500, 4000, 5000, 
6000, and 8000 K.
\end{itemize}

\begin{figure}
\resizebox{\hsize}{!}{\includegraphics{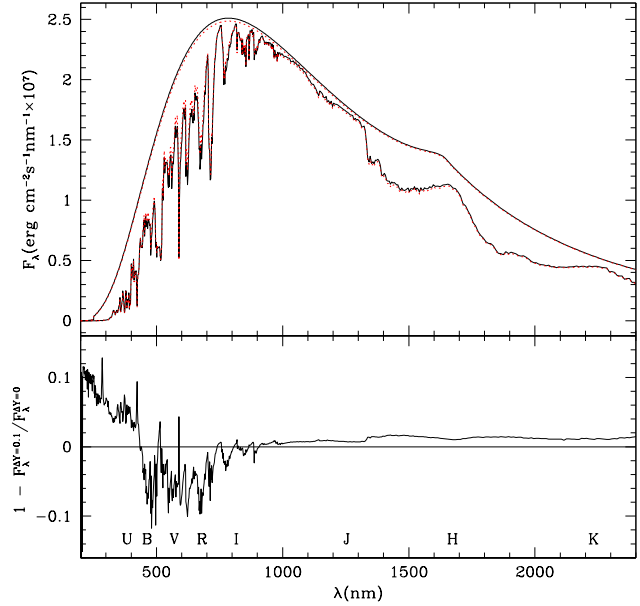}}
\caption{Top panel: Spectral energy distribution for a star with 
$\Teff=3500$~K, $\log g=4.5$, and $\mh=-0.5$, for both the $\Delta
Y=0$ and $\Delta Y=0.1$ cases (full and dotted lines,
respectively). The two upper lines compare the continua, the two lower
lines compare the emergent fluxes ($F_\lambda^{\Delta Y=0.1}$ and
$F_\lambda^{\Delta Y=0}$). Bottom panel: the relative difference
between the above spectral energy distributions, illustrated by means
of $1-F_\lambda^{\Delta Y=0.1}/F_\lambda^{\Delta Y=0}$. The plot also
indicates the approximate location of Johnson-Cousins $UBVRIJHK$
pass-bands.}
\label{fig_spectra} 
\end{figure}

As an example, Figure~\ref{fig_spectra} compares spectral energy
distributions, differing only for the He content, for a relatively
cool dwarf of intermediate metallicity. In the top panel, the upper
continous lines indicate the emergent flux due to the only continous
opacities, while the lower lines are the emergent flux due to both
continous and line opacities. The He-enhancement has a modest impact
on the emergent spectra. This is evident in the bottom panel of
Fig.~\ref{fig_spectra}, where the quantity $1-F_\lambda^{\Delta
Y=0.1}/F_\lambda^{\Delta Y=0}$ is plotted.  {The differences
between the two spectra amount to just a few percent, which translate
in maximum changes of just a few hundredths of magnitude in bolometric
corrections (see Sect.~\ref{sec_synphot} below).}

\begin{figure*}
\resizebox{0.7\hsize}{!}{\includegraphics{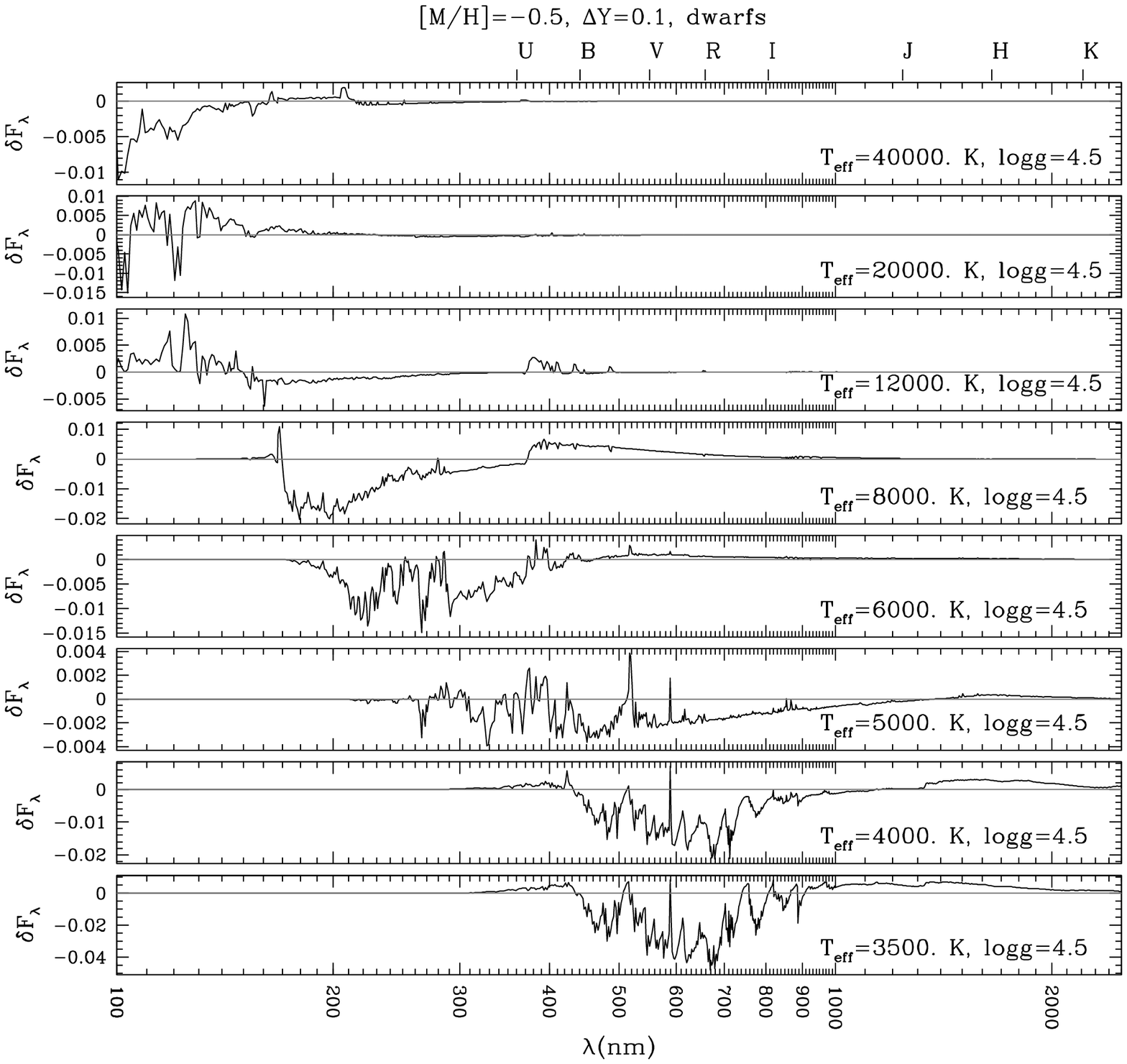}}
\\
\resizebox{0.7\hsize}{!}{\includegraphics{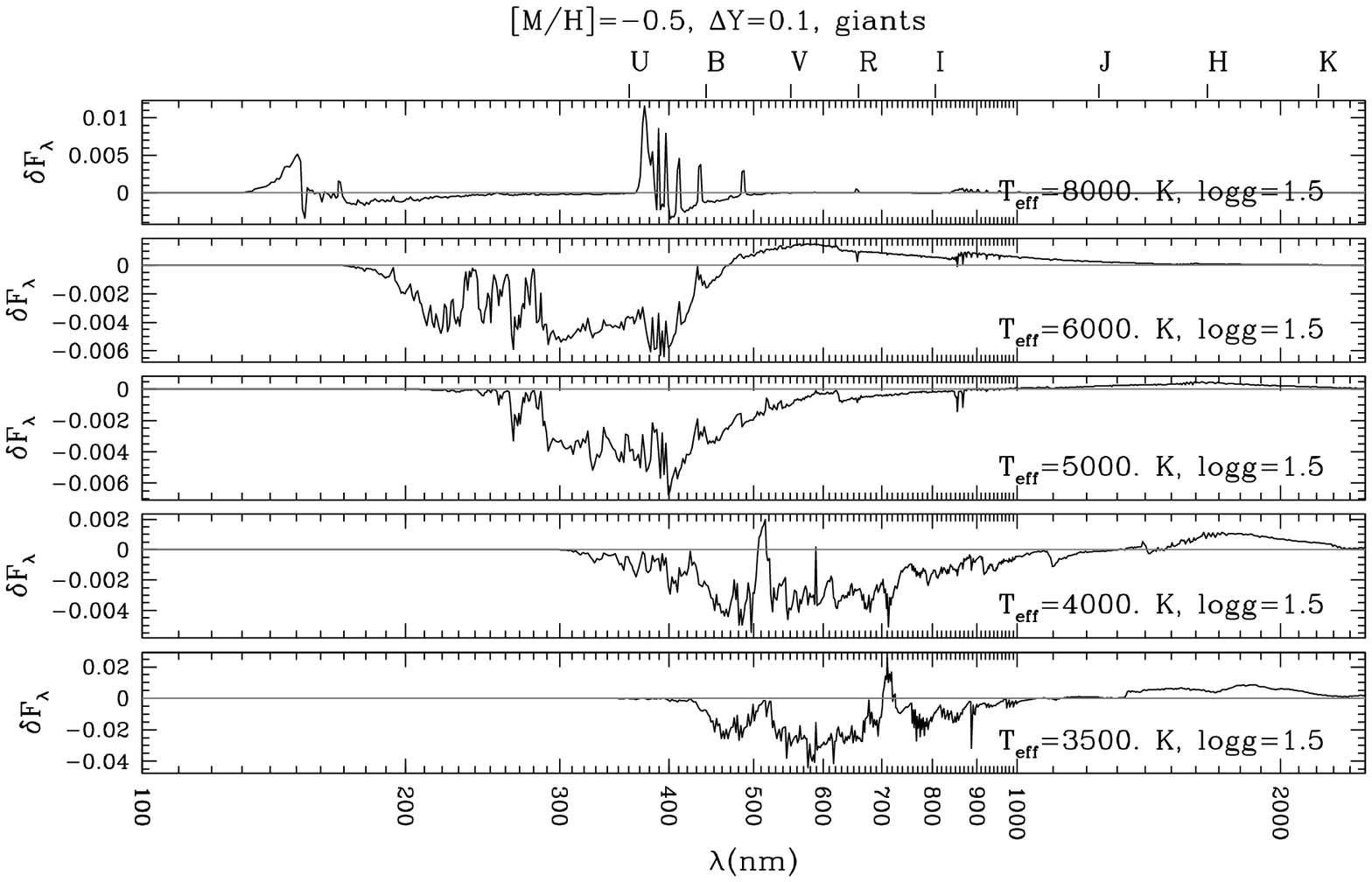}}
\caption{The quantity $\delta F_\lambda$ (see eq.~\protect\ref{eq_deltaf}) 
as a function of $\lambda$, as derived for the $\mh=-0.5$ spectra with
$\Delta Y=0$ and $\Delta Y=0.1$. The {top and bottom} panels show
sequences of decreasing \Teff\ for both dwarfs
and giants, respectively. Similar figures are available also for
$\mh=-1.5$ and $\mh=+0.5$, and can be provided upon request. }
\label{fig_diffspectra} 
\end{figure*}

{Moreover, some of the differences seen in the bottom panel of
Fig.~\ref{fig_spectra} are of no concern because they appear at
spectral regions where the emergent flux is very small (for instance,
for $\lambda<400$~nm in the figure)}.  In order to better illustrate
the differences in the computed spectra which are due only to the
variation in the He content, Fig.~\ref{fig_diffspectra} presents a
complete series of plots of the quantity $\delta F_\lambda$, defined
as
\begin{equation}
\delta F_\lambda = \frac{ F_\lambda^{\Delta Y=0} - 
	F_\lambda^{\Delta Y=0.1} }
	{F_{{\rm max}}^{\Delta Y=0}}
\label{eq_deltaf}
\end{equation}
where {$F_{{\rm max}}^{\Delta Y=0}$ is the maximum flux of
the $F_{\lambda}^{\Delta Y=0}$ spectrum}. By plotting the quantity
$\delta F_\lambda$, we evidence only the differences that occur in the
spectral region which is more relevant in terms of flux.  {This
allows a quick evaluation of the changes that are potentially more
important to the photometry}. Of course, differences between the
$\Delta Y=0$ and $\Delta Y>0$ cases occur over the complete range of
$\lambda$.

\section{Synthetic photometry and results}
\label{sec_synphot}

We have performed synthetic photometry for the above-mentioned energy
distributions using the same formalism as in Bessell et al. (1998)
and Girardi et al. (2002). Since we are just interested in the changes
that the enhanced He can have in the synthetic photometry, the
equation to be used is:
\begin{equation}
\Delta BC_{S_\lambda}  =  -2.5\,\log\left(
	\frac { \int_{\lambda_1}^{\lambda_2} 
		\lambda F_{\lambda}^{\Delta Y>0} \, 
		S_\lambda \diff\lambda }{ \int_{\lambda_1}^{\lambda_2} 
		\lambda F_{\lambda}^{\Delta Y=0} \, 
		S_\lambda \diff\lambda } \right)
\label{eq_deltabc} 
\end{equation}
where $S_\lambda$ is the total throughput in the filter under
consideration, defined in the interval $[\lambda_1,\lambda_2]$.  These
$\Delta BC_{S_\lambda}$ directly tell us the effect of He-enhancement
on the absolute magnitudes. The effect on colours can be simply
derived by the differences in $\Delta BC_{S_\lambda}$ for two filters.

\begin{figure}
\resizebox{\hsize}{!}{\includegraphics{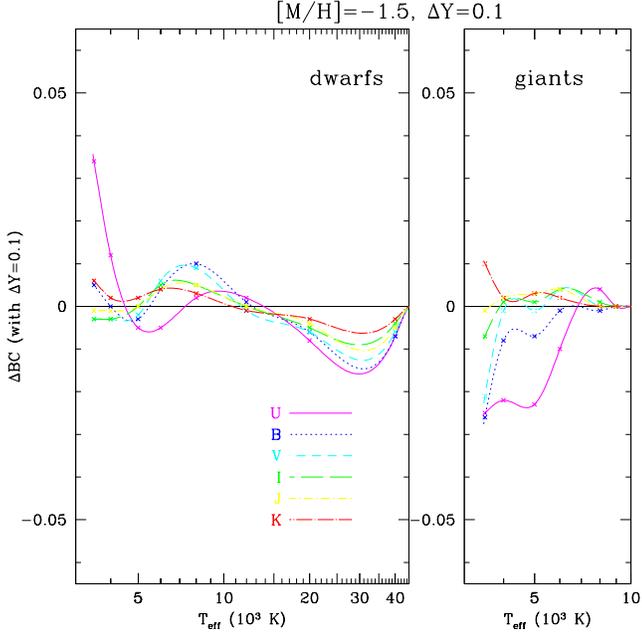}}
\caption{The $\Delta BC_{S_\lambda}$ quantities as a function of \Teff, 
for both dwarfs and giants of metallicity $\mh=-1.5$ and $\Delta
Y=0.1$, and for some of the Johnson-Cousins-Glass $UBVRIJHK$
filters. The small crosses are the $\Delta BC_{S_\lambda}$ values
effectively computed in this work; they are linked by natural spline
curves just for the sake of a better distinction between the different
filters.}
\label{fig_diff_m15}
\end{figure}

\begin{figure}
\resizebox{\hsize}{!}{\includegraphics{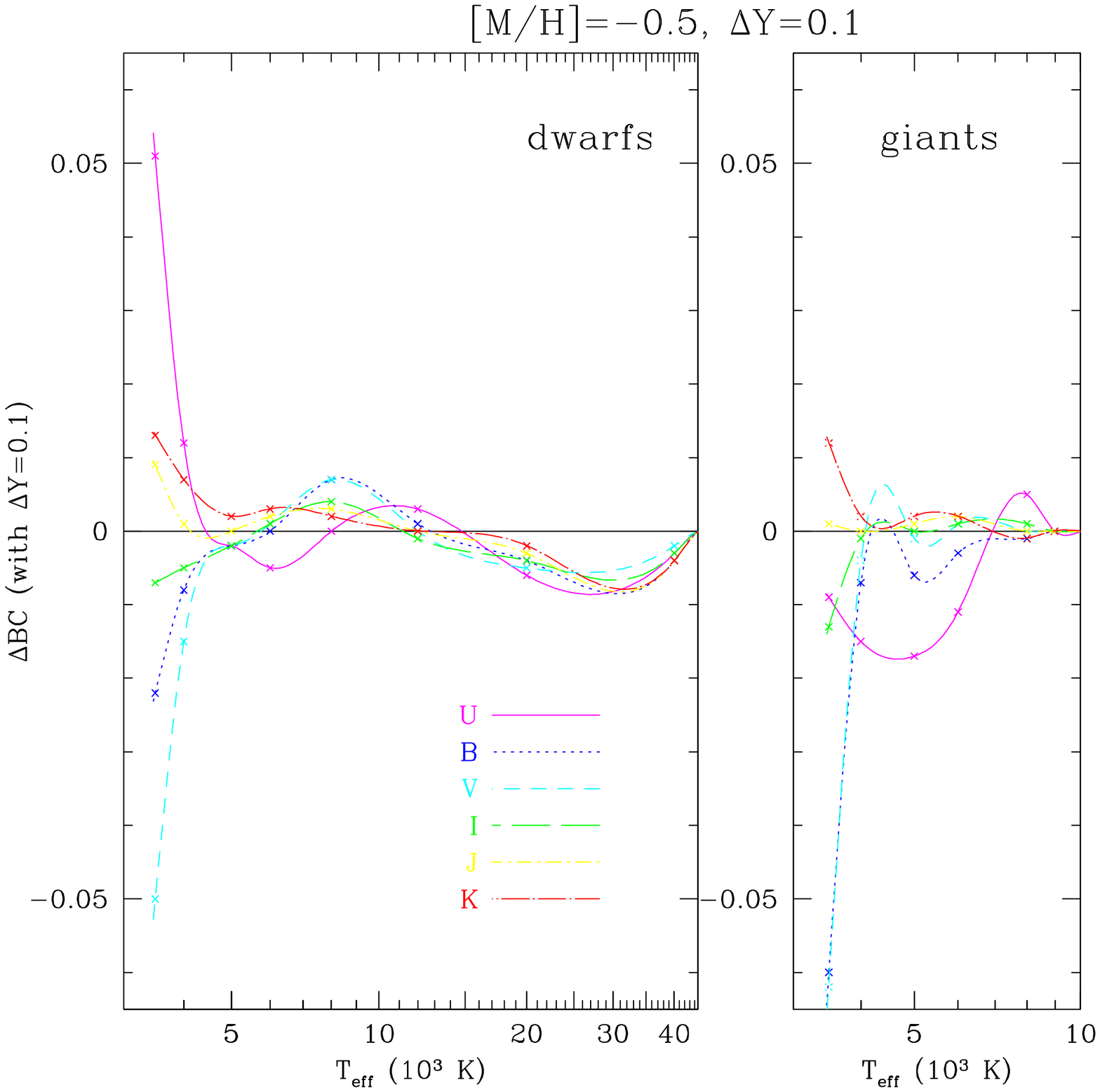}}
\caption{The same as Fig.~\protect\ref{fig_diff_m15}, but for 
$\mh=-0.5$ and $\Delta Y=0.1$.}
\label{fig_diff_m05}
\end{figure}

\begin{figure}
\resizebox{\hsize}{!}{\includegraphics{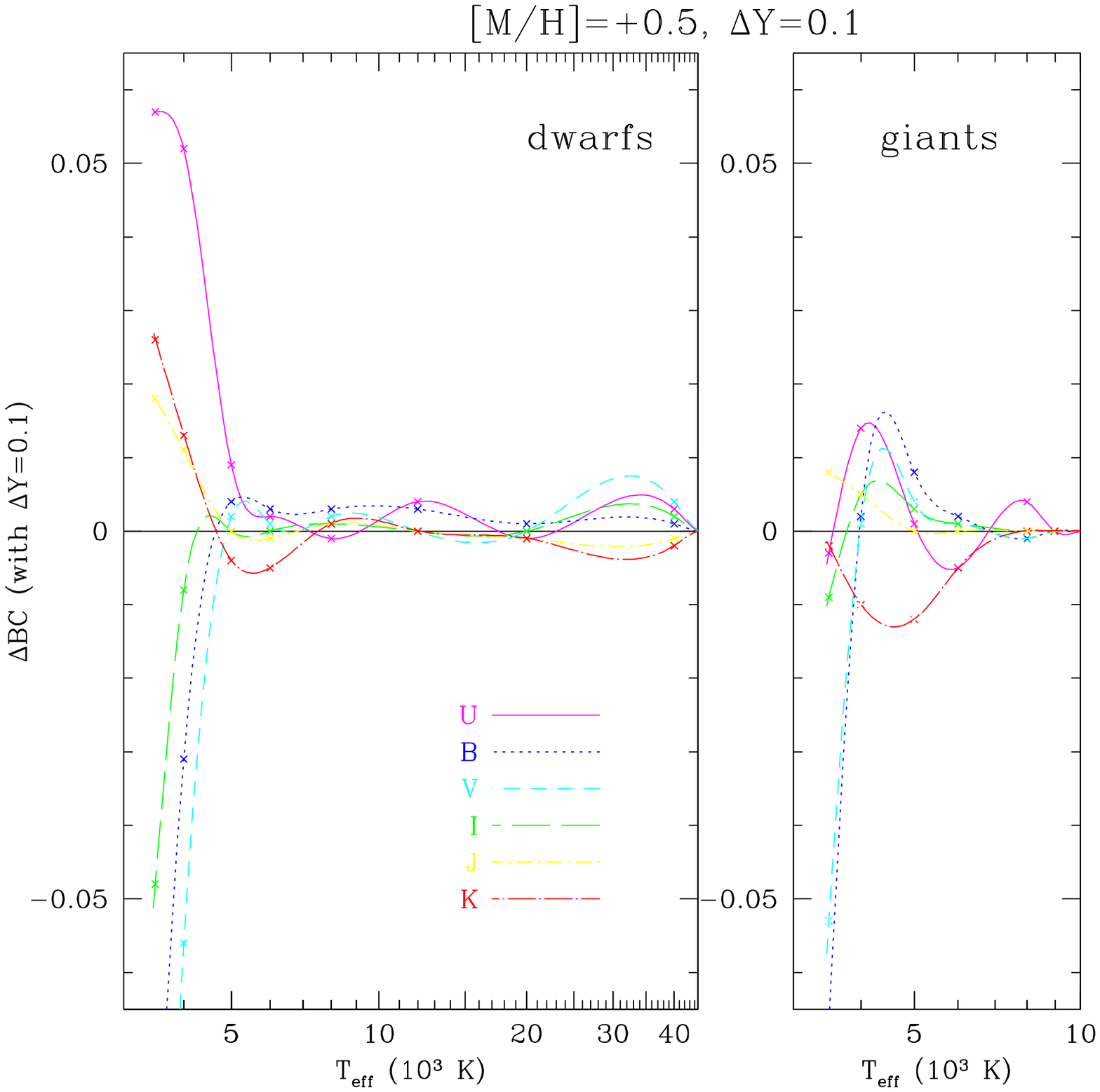}}
\caption{The same as Fig.~\protect\ref{fig_diff_m15}, but for 
$\mh=+0.5$ and $\Delta Y=0.1$.}
\label{fig_diff_p05}
\end{figure}

\begin{figure}
\resizebox{\hsize}{!}{\includegraphics{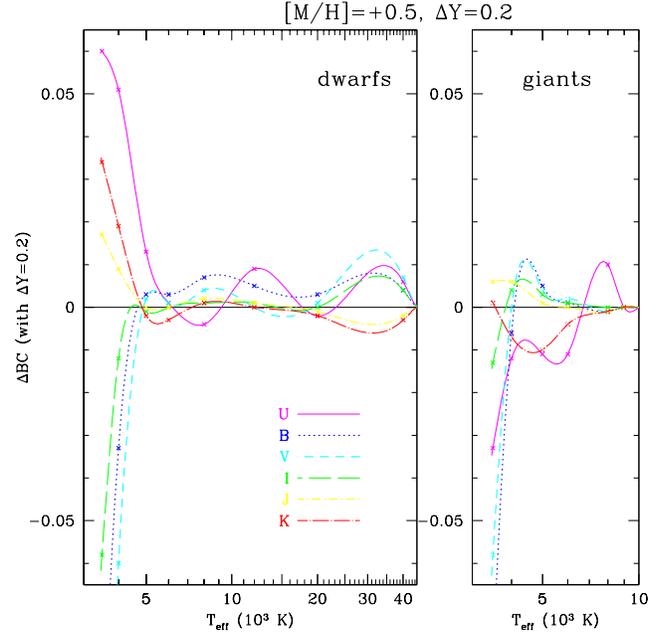}}
\caption{The same as Fig.~\protect\ref{fig_diff_m15}, but for 
$\mh=+0.5$ and $\Delta Y = 0.2.$  }
\label{fig_diff_p02p05}
\end{figure}

Figures \ref{fig_diff_m15} to \ref{fig_diff_p02p05} illustrate the
behaviour of $\Delta BC_{S_\lambda}$ as a function of \Teff, for both
dwarfs and giants, for all \mh\ and $\Delta Y$ values considered in
this work, and for the specific case of Johnson-Cousins-Glass
$UBVRIJHK$ filters. The filter curves were taken from Bessell (1990)
and Bessell \& Brett (1988). The same data are tabulated in Table~1,
and is provided in electronic form at {\tt
http://pleiadi.oapd.inaf.it}.

\section{Discussion and conclusions}
\label{sec_conclu}

As can be readily seen in Figs.~\ref{fig_diff_m15} to
\ref{fig_diff_p02p05}, the effect of He-enhancement in the BC is overall
quite modest. The most remarkable result for all cases considered here
is that {\em $|\Delta BC_{S_\lambda}|$ are smaller than 0.009~mag for
all stars with $\Teff \ge 5000$~K and for all pass-bands redder than
$U$.} The typical $|\Delta BC_{S_\lambda}|$ values in these cases are
even smaller, of the order of 0.005~mag. In general the corresponding
shifts in absolute magnitude are well below the typical errors in
photometric observations. Since in most cases the $\Delta
BC_{S_\lambda}$ behave in a similar way for different filters, the
effects in colours are of even smaller magnitude. It is clear to us
that the effect of He-enhancement is small enough to be neglected in
these cases.

Significant values of $|\Delta BC_{S_\lambda}|$ are met just in a few
situations; namely for the $U$ filter and at intermediate values of
\Teff\, i.e. between 4000 and 8000~K, $|\Delta BC_{S_\lambda}|$ become
slightly higher but anyway still of the order of 0.02~mag. This is
already an effect that could be detected in the (as far as we know,
rare) case of high-precision photometry in the $U$ passband.
The considered range of \Teff\ is high enough to include the turn-off
region of metal-poor globular clusters, and part of their horizontal
branch. Therefore, in very specific cases the effect of He-enhancement
may have to be considered in globular clusters.

On the other hand for \Teff\ approaching the value of 3500~K and for
most filters in the blue part of the spectrum (from $U$ to $V$),
$|\Delta BC_{S_\lambda}|$ become significantly larger and can amount
to as much as 0.06~mag at $\mh \leq -0.5$, and 0.15~mag at $\mh =
+0.5$.  These low \Teff\ values are those typical of early-M giants,
including for instance the tip of the RGB (TRGB) at {old ages and
moderately low metallicities ($\feh\sim-0.7$)}. Fortunately in the
same range of low \Teff\ the red and infrared passbands present small
$\Delta BC_S$ corrections, in particular in the $I$ band $\Delta
BC_{I}$ becomes smaller than $0.015$~mag. {We notice that distance
determinations of resolved galaxies via the TRGB $I$-band magnitude
should not be affected by possible galaxy-to-galaxy changes in the
mean He content, since they usually refer to stars hotter than
$\sim4000$~K, for which the possible $\Delta BC_{I}$ corrections are
even smaller than at 3500~K.}

However, we remark that it is not at all clear whether the significant
values of $|\Delta BC_{S_\lambda}|$ at $\Teff\sim3500$~K are a serious
problem owing to the well-known uncertainties of the ATLAS9 models for
$\Teff\leq4000$~K.  For instance, at $\Teff\sim3500$~K starts the
formation of strong molecular bands in the stellar spectra, which are
not accurately reproduced -- at least not at the level of a few
percent -- by present-day ATLAS9 models (see for instance Fluks et
al. 1994).  Among the reasons there is the lack in the line opacity
computations of both triatomic molecules (with exception for H$_{2}$O
which is considered) and of numerous diatomic molecular transitions.

Therefore, the significant changes in $\Delta BC_{S_\lambda}$ that we
find at low \Teff\ may be just one additional -- and secondary --
problem in a field that is already complicated in itself, and for
which synthetic photometry has always been recognized not to provide
accurate answers.

In conclusion, we find that the effects of changes in He abundances
among stellar populations are quite modest when we look at the stellar
atmospheres and their predicted bolometric corrections. Therefore, the
use of tables of BCs computed for a single $Y(Z)$ relation, is an
acceptable approximation in most cases. We provide tables for $\Delta
BC_{S_\lambda}$ in a series of \mh, $\Delta Y$, and \Teff\ values,
that may help the reader to evaluate whether this is an issue in the
interpretation of their observations.  The effects of changing $Y$ by
as much as 0.1, instead, may have a quite high impact on the stellar
evolutionary tracks, and have to be considered whenever it is
suspected, as in the case of $\omega$~Cen.

\begin{acknowledgements}
This work is funded by the grant INAF PRIN/05 1.06.08.03 ``A
theoretical lab for stellar population studies''.
\end{acknowledgements}


\section*{References}

\begin{description}

\item Bertelli, G., et al., 1994, A\&AS, 106, 275
\item Bessell, M.S., 1990, PASP, 102, 1181
\item Bessell, M.S., \& Brett J.M., 1988, PASP, 100, 1134
\item Bessell, M.S., Castelli F., \& Plez B., 1998, A\&A. 333, 231
\item Caloi, V., \& D'Antona, F., 2005, A\&A, 435, 987
\item Castelli, F., 2005, MSAIS, 8, 34 
\item Castelli, F., Gratton, R.G., \& Kurucz R.L., 1997, A\&A, 318, 841
\item Castelli, F., \& Kurucz, R.~L., 2003, IAU Symposium, 210, 20P
\item D'Antona, F. et al., 2005 ApJ, 631, 868 
\item Fluks, M.A, Plez, B., The, P.S., et al., 1994, A\&AS, 105, 311
\item Girardi, L., Bressan, A., Bertelli, G., Chiosi, C., 2000, A\&AS, 141,371
\item Grevesse, N., \& Sauval, A. J., 1998, SSR 85, 161
\item Jimenez, R. et al., 2003, Science, 299, 1552
\item Kurucz, R.L., 1993, in  IAU
     	Symp. 149: The Stellar Populations of Galaxies, 
	eds. B. Barbuy, A. Renzini, Dordrecht, Kluwer, p. 225
\item Kurucz, R. L,. 2005, MSAIS, 8, 14 
\item Lebreton, Y. et al.., 2001, A\&A, 374, 540
\item Lee, Y.-W., et al., 2005, ApJ, 621, L57
\item Pagel, B.E.J., \& Portinari, L., 1998, MNRAS,298,747
\item Peimbert, M. et al. ., 2000, ApJ, 541, 688
\item Piotto, G., et al., 2005, ApJ, 621, 777
\item Spergel, D.N., et al., 2003, ApJS, 148, 175
\item Spergel, D.N., et al., 2006, astro-ph/0603449

\end{description}  

\begin{longtable}{rr|rrrrrrrrr}
\caption{
\label{tab_bc}
$\Delta BC_{S_\lambda}$ values (in mag) for the $UBVRIJHK$ system of
Bessell (1990) and Bessell \& Brett (1988).  }
\\
\hline\hline
\Teff & $\log g$ & $U$ & $BX$ & $B$ & $V$ & $R$ & $I$ & $J$ & $H$ & $K$ \\
\hline
\endfirsthead
\caption{continued.}\\
\hline\hline
\Teff & $\log g$ & $U$ & $BX$ & $B$ & $V$ & $R$ & $I$ & $J$ & $H$ & $K$ \\
\hline
\endhead
\hline
\endfoot
\multicolumn{11}{c}{$\mh=-1.5,\Delta Y=0.1$} \\ 
3500 & 4.50 & 0.034 & 0.005 & 0.005 & -0.003 & -0.006 & -0.003 & -0.001 & 0.000 & 0.006 \\ 
4000 & 4.50 & 0.012 & 0.000 & 0.000 & -0.003 & -0.004 & -0.003 & -0.001 & 0.001 & 0.002 \\
5000 & 4.50 & -0.005 & -0.003 & -0.003 & -0.002 & -0.001 & 0.000 & 0.000 & 0.002 & 0.002 \\ 
6000 & 4.50 & -0.005 & 0.004 & 0.004 & 0.006 & 0.005 & 0.005 & 0.004 & 0.005 & 0.004 \\ 
8000 & 4.50 & 0.002 & 0.010 & 0.010 & 0.009 & 0.007 & 0.005 & 0.005 & 0.003 & 0.003 \\ 
12000 & 4.50 & 0.002 & 0.001 & 0.001 & -0.001 & 0.000 & 0.000 & 0.000 & 0.000 & -0.001 \\ 
20000 & 4.50 & -0.008 & -0.006 & -0.006 & -0.006 & -0.005 & -0.005 & -0.004 & -0.003 & -0.003 \\ 
40000 & 4.50 & -0.007 & -0.007 & -0.007 & -0.006 & -0.004 & -0.004 & -0.005 & -0.005 & -0.003 \\ 
3500 & 1.50 & -0.025 & -0.027 & -0.026 & -0.021 & -0.014 & -0.007 & -0.001 & 0.009 & 0.010 \\ 
4000 & 1.50 & -0.022 & -0.008 & -0.008 & -0.001 & 0.000 & 0.001 & 0.002 & 0.001 & 0.002 \\ 
5000 & 1.50 & -0.023 & -0.007 & -0.007 & -0.001 & 0.002 & 0.001 & 0.003 & 0.004 & 0.003 \\ 
6000 & 1.50 & -0.010 & -0.001 & -0.001 & 0.004 & 0.004 & 0.004 & 0.004 & 0.002 & 0.002 \\ 
8000 & 1.50 & 0.004 & -0.001 & -0.001 & 0.000 & 0.000 & 0.001 & 0.000 & -0.001 & 0.000 \\ 
\hline
\multicolumn{11}{c}{$\mh=-0.5,\Delta Y=0.1$} \\ 
3500 & 4.50 & 0.051 & -0.024 & -0.022 & -0.050 & -0.044 & -0.007 & 0.009 & 0.014 & 0.013 \\ 
4000 & 4.50 & 0.012 & -0.008 & -0.008 & -0.015 & -0.014 & -0.005 & 0.001 & 0.008 & 0.007 \\ 
5000 & 4.50 & -0.002 & -0.002 & -0.002 & -0.002 & -0.002 & -0.002 & 0.000 & 0.002 & 0.002 \\ 
12000 & 4.50 & 0.003 & 0.001 & 0.001 & 0.000 & 0.000 & -0.001 & 0.000 & 0.000 & 0.000 \\ 
20000 & 4.50 & -0.006 & -0.005 & -0.004 & -0.005 & -0.004 & -0.004 & -0.003 & -0.003 & -0.002 \\ 
40000 & 4.50 & -0.003 & -0.004 & -0.004 & -0.002 & -0.002 & -0.003 & -0.004 & -0.004 & -0.004 \\ 
3500 & 1.50 & -0.009 & -0.061 & -0.060 & -0.062 & -0.037 & -0.013 & 0.001 & 0.009 & 0.012 \\ 
4000 & 1.50 & -0.015 & -0.007 & -0.007 & -0.004 & -0.003 & -0.001 & 0.000 & 0.002 & 0.002 \\ 
5000 & 1.50 & -0.017 & -0.005 & -0.006 & -0.001 & 0.000 & 0.000 & 0.001 & 0.002 & 0.002 \\ 
6000 & 1.50 & -0.011 & -0.002 & -0.003 & 0.001 & 0.002 & 0.001 & 0.002 & 0.001 & 0.002 \\ 
8000 & 1.50 & 0.005 & -0.001 & -0.001 & 0.000 & 0.000 & 0.001 & 0.000 & 0.000 & -0.001 \\ 
\hline
\multicolumn{11}{c}{$\mh=+0.5,\Delta Y=0.1$} \\ 
3500 & 4.50 & 0.057 & -0.096 & -0.092 & -0.130 & -0.088 & -0.048 & 0.018 & 0.034 & 0.026 \\ 
4000 & 4.50 & 0.052 & -0.032 & -0.031 & -0.056 & -0.047 & -0.008 & 0.011 & 0.013 & 0.013 \\ 
5000 & 4.50 & 0.009 & 0.004 & 0.004 & 0.002 & 0.001 & 0.000 & 0.000 & -0.004 & -0.004 \\ 
6000 & 4.50 & 0.002 & 0.003 & 0.003 & 0.001 & 0.000 & 0.000 & -0.001 & -0.004 & -0.005 \\ 
8000 & 4.50 & -0.001 & 0.003 & 0.003 & 0.002 & 0.002 & 0.001 & 0.001 & 0.000 & 0.001 \\ 
12000 & 4.50 & 0.004 & 0.002 & 0.003 & 0.000 & 0.000 & 0.000 & 0.000 & 0.001 & 0.000 \\ 
20000 & 4.50 & -0.001 & 0.001 & 0.001 & 0.000 & 0.000 & 0.000 & -0.001 & -0.001 & -0.001 \\ 
40000 & 4.50 & 0.003 & 0.001 & 0.001 & 0.004 & 0.003 & 0.002 & -0.001 & -0.002 & -0.002 \\ 
3500 & 1.50 & -0.003 & -0.067 & -0.067 & -0.053 & -0.010 & -0.009 & 0.008 & -0.002 & -0.002 \\ 
4000 & 1.50 & 0.014 & 0.002 & 0.002 & 0.001 & 0.003 & 0.005 & 0.005 & -0.008 & -0.010 \\ 
5000 & 1.50 & 0.001 & 0.008 & 0.008 & 0.004 & 0.004 & 0.003 & 0.000 & -0.009 & -0.012 \\ 
6000 & 1.50 & -0.005 & 0.002 & 0.002 & 0.001 & 0.002 & 0.001 & 0.000 & -0.004 & -0.005 \\ 
8000 & 1.50 & 0.004 & -0.001 & -0.001 & -0.001 & 0.000 & 0.000 & 0.000 & 0.000 & 0.000 \\ 
\hline
\multicolumn{11}{c}{$\mh=+0.5,\Delta Y=0.2$} \\ 
3500 & 4.50 & 0.060 & -0.107 & -0.103 & -0.147 & -0.102 & -0.058 & 0.017 & 0.042 & 0.034 \\ 
4000 & 4.50 & 0.051 & -0.035 & -0.033 & -0.060 & -0.052 & -0.012 & 0.009 & 0.019 & 0.019 \\ 
5000 & 4.50 & 0.013 & 0.004 & 0.003 & 0.002 & -0.001 & -0.001 & 0.000 & -0.002 & -0.002 \\ 
6000 & 4.50 & 0.001 & 0.003 & 0.003 & 0.001 & 0.000 & 0.000 & 0.000 & -0.002 & -0.003 \\ 
8000 & 4.50 & -0.004 & 0.006 & 0.007 & 0.004 & 0.003 & 0.001 & 0.002 & 0.000 & 0.001 \\ 
12000 & 4.50 & 0.009 & 0.005 & 0.005 & 0.000 & 0.001 & 0.001 & 0.001 & 0.002 & 0.000 \\ 
20000 & 4.50 & -0.002 & 0.003 & 0.003 & 0.001 & 0.000 & 0.000 & -0.001 & -0.001 & -0.002 \\ 
40000 & 4.50 & 0.006 & 0.004 & 0.004 & 0.007 & 0.006 & 0.004 & -0.002 & -0.003 & -0.003 \\ 
3500 & 1.50 & -0.033 & -0.081 & -0.081 & -0.058 & -0.014 & -0.013 & 0.006 & -0.002 & 0.001 \\ 
4000 & 1.50 & -0.012 & -0.006 & -0.006 & 0.000 & 0.002 & 0.004 & 0.006 & -0.007 & -0.007 \\ 
5000 & 1.50 & -0.011 & 0.005 & 0.005 & 0.004 & 0.004 & 0.003 & 0.001 & -0.007 & -0.010 \\ 
6000 & 1.50 & -0.011 & 0.002 & 0.001 & 0.002 & 0.002 & 0.001 & 0.000 & -0.003 & -0.004 \\ 
8000 & 1.50 & 0.010 & -0.001 & -0.001 & -0.001 & 0.000 & 0.000 & -0.001 & -0.001 & -0.001 \\ 

\end{longtable}

\end{document}